\documentclass[12pt]{revtex4}

\usepackage{amsmath}
\usepackage{graphicx}

\newcommand{\defn}{\textit}

\newcommand{\eref}[1]{(\ref{#1})}
\newcommand{\av}[1]{\langle#1\rangle}

\newcommand{\matA}{\mathbf{A}}

\newcommand{\matAt}{\mathbf{A}^{\mathrm{T}}}
\newcommand{\lmax}{\lambda_{\mathrm{max}}}
\newcommand{\kin}{k^{\mathrm{in}}}
\newcommand{\kout}{k^{\mathrm{out}}}

\newcommand{\vecw}{\mathbf{w}}
\newcommand{\vecl}{\mathbf{l}}

\newlength{\figurewidth}
\ifdim\columnwidth<20cm
\else
\fi
\begin{document}
\title{A network-based ranking system for US college football}
\author{Juyong Park}
\affiliation{Department of Physics and Center for the Study of Complex
Systems, University of Michigan, Ann Arbor, MI 48109}
\author{M.~E.~J. Newman}
\affiliation{Department of Physics and Center for the Study of Complex
Systems, University of Michigan, Ann Arbor, MI 48109}
\begin{abstract}
American college football faces a conflict created by the desire to stage
national championship games between the best teams of a season when there
is no conventional playoff system to decide which those teams are.
Instead, ranking of teams is based on their record of wins and losses
during the season, but each team plays only a small fraction of eligible
opponents, making the system underdetermined or contradictory or both.  It
is an interesting challenge to create a ranking system that at once is
mathematically well-founded, gives results in general accord with received
wisdom concerning the relative strengths of the teams, and is based upon
intuitive principles, allowing it to be accepted readily by fans and
experts alike.  Here we introduce a one-parameter ranking method that
satisfies all of these requirements and is based on a network
representation of college football schedules.
\end{abstract}
\maketitle

\section{Introduction}
Inter-university competition in (American) football is big business in the
United States.  Games are televised on national TV; audiences number in the
millions and advertising revenues in the hundreds of millions (of US
dollars).  Strangely, however, there is no official national championship
in college football, despite loudly-voiced public demand for such a thing.
In other sports, such as soccer or basketball, there are knockout
competitions in which schedules of games are drawn up in such a way that at
the end of the competition there is an undisputed ``best'' team---the only
team in the league that remains unbeaten.  A simple pairwise elimination
tournament is the most common scheme.

The difficulty with college football is that games are mostly played in
\defn{conferences}, which are groups of a dozen or so colleges chosen on
roughly geographic grounds.  In a typical season about 75\% of games are
played between teams belonging to the same conference.  As a result there
is normally an undisputed champion for each individual conference, but not
enough games are played between conferences to allow an overall champion to
be chosen unambiguously.  Some other sports also use the conference system,
and in those sports an overall champion is usually chosen via a separate
knockout tournament organized among the winners and runners up in the
individual conferences.  In college football, however, for historical and
other reasons, there is no such post-season tournament.

To fulfill the wishes of the fans for a national championship, therefore,
several of the major conferences have adopted a system called the Bowl
Championship Series (BCS, \mbox{\texttt{www.bcsfootball.org}}), in which
one of four existing post-season ``bowl games''---the Rose, Sugar, Fiesta,
and Orange Bowls---is designated the national championship game on a
rotating basis and is supposed to match the top two teams of the regular
season~\cite{Dunnavant:2004ns}.  (For the 2004 season it was the Orange
Bowl; in the upcoming 2005 season it will be the Rose Bowl.)  The problem
is how to decide which the top teams are.  One can immediately imagine many
difficulties.  Simply choosing unbeaten teams will not necessarily work:
what if there are more than two, or only one, or none?  How should one
account for teams that play different numbers of regular-season games, and
for ``strength of schedule''---the fact that some teams by chance
inevitably play against tougher opponents than others?  What about margins
of victory?  Should a decisive victory against your opponent count for more
than a narrow victory?  Should home games be counted differently from away
games?

The problem of ranking competitors based on an incomplete set of pairwise
comparisons is a well-studied one, both in football and other sports, and
more generally~\cite{David:1988ri}.  Many different methods and algorithms
have been proposed, frequently taking into account the issues mentioned
above~\cite{Keener:1993jj,Callagan:2004mh,Harville:2003uj,Stefani:1997bb,Glickmann:1998mz,Marcus:2001az}.
Currently football teams are ranked using a weighted composite score called
the BCS ranking that combines a number of these methods with polls of
knowledge human judges.  The formula used changes slightly from year to
year; the most recent version averages six computer algorithms~\footnote{A
comprehensive summary and comparisons of known computer ranking methods for
American college football, including the
\emph{Daniel ranking} system which is similar to, through more rudimentary
than, the method presented here, can be found at
\mbox{\texttt{http://homepages.cae.wisc.edu/\~{}dwilson/rsfc/rsfc.shtml}}}
and two human polls. (One of them, the Associated Press (AP) poll, has opted
out of the system starting from the 2005 season.)  There is, however,
considerable unhappiness about the system and widespread disagreement about
how it should be improved~\footnote{It was originally hoped that the BCS
rankings would help generate consensus about the true number 1 and 2 teams,
resulting in an undisputed national champion, but it hasn't always worked
out that way.  Most recently in 2003, for instance, the AP poll awarded its
top spot to the University of Southern California, contradicting the
overall BCS rankings, which awarded top honors to the Louisiana State
University, and resulting in a ``split'' national title of the kind the
system was designed to avoid.}.  There is, thus, plenty of room for
innovation.

In this paper we present a new method of ranking based on a mathematical
formulation that corresponds closely to the types of arguments typically
made by sports fans in comparing teams.  Our method turns out to be
equivalent to a well-known type of centrality measure defined on a directed
network representing the pattern of wins and losses in regular-season
games.

\section{Definition of the method}
Perhaps the simplest measure of team standing is the win-loss differential,
i.e.,~the number of games a team wins during the season minus the number it
loses.  (In American football there are no tied games---games are extended
until there is a winner.)  Indeed, the win-loss differential is almost the
only measure that everyone seems to agree upon.  It is unfortunate
therefore that in practice it correlates rather poorly with expert opinions
about which teams are best, for many of the reasons cited in the previous
section, such as variation in strength of schedule.  As we show here,
however, we can correct for these problems, at least in part, by
considering not just direct wins and losses, but also indirect ones.

\begin{figure}
\includegraphics[width=40mm]{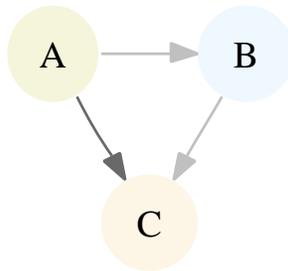}
\caption{If team~A has beaten team~B, and team~B has beaten team~C, team~A
scores an indirect win over team~C (indicated by the bold arrow).}
\label{indirect}
\end{figure}

One often hears from sports fans arguments of the form: ``Although my
team~A didn't play your team~C this season, it did beat~B who in turn
beat~C.  Therefore A is better than C and would have won had they played a
game.''  (See Fig.~\ref{indirect}.)  In fact, the argument is usually
articulated with less clarity than this and more beer, but nonetheless we
feel that the general line of reasoning has merit.  What the fan is saying
is that, in addition to a real, physical win (loss) against an opponent, an
\emph{indirect win (loss)} of the type described should also be considered
indicative of a team's strength (weakness).  It is on precisely this kind
of reasoning that we base our method of ranking.

\subsection{The college football schedule as a network}
\begin{figure}
\includegraphics[width=120mm]{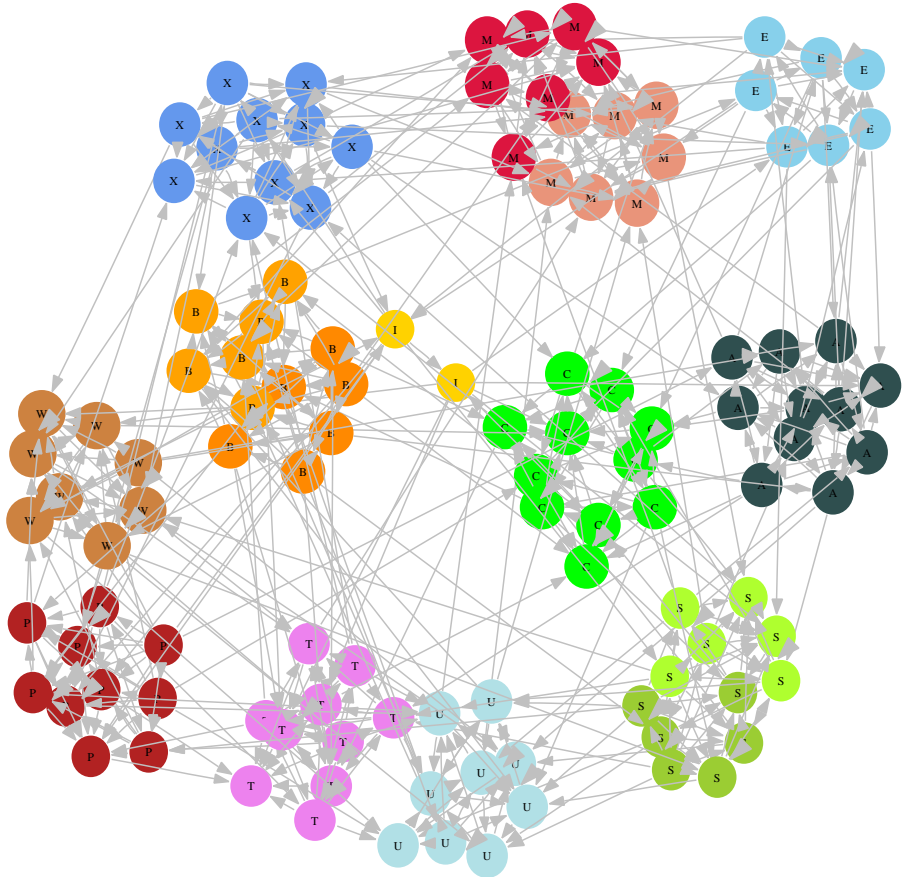}
\caption{A graphical representation of the regular season schedule of
division I-A teams in 2004.  Teams are divided up by conference (A =
Atlantic Coast, E = Big East, X = Big Ten, B = Big XII, C = Conference USA,
M = Mid-American, P = Pac Ten, W = Mountain West, S = Southeastern, U = Sun
Belt, T = Western Athletic, I = Independent).  Directed edges point from
winners to losers.}
\label{2004outcome}
\end{figure}

The schedule of games for a season can be represented as a network or graph
in which the vertices represent colleges and there is an edge between two
colleges if they played a regular-season game during the season of
interest~\cite{GN02}.  Furthermore, we can represent the winner and loser
of each game by making the network directed.  We place an arrow on each
edge pointing from the winner of the corresponding game to the loser.  An
example of such a network, for the 2004 season, is shown in
Fig.\ref{2004outcome}.  (The direction of the arrows is a matter of
convention; we could have made the opposite choice had we wished and the
network would still contain the same information.)

Direct losses and wins of a team in this network correspond to edges
running directly to and from that team, and indirect losses and wins, as
defined above, correspond to \emph{directed paths of length~2} in the
network, to and from the team.

A particularly nice property of these indirect wins is that a direct win
against a strong opponent---a team that has itself won many games---is
highly rewarding, giving you automatically a large number of indirect wins.
Thus, when measured in terms of indirect wins, the ranking of a team
automatically allows for the strength of schedule.

And there is no need to stop here: one can consider higher-order indirect
wins (or losses) of the form A beats B beats C beats D, and so forth.
These correspond to directed paths in the network of length three or more.
Our proposed ranking scheme counts indirect wins and losses at all
distances in the network, but those at greater distances count for less,
because we feel it natural that a direct win against a team should count
for more than the mere supposed victory of an indirect win.

Mathematically, we can express these ideas in terms of the adjacency
matrix~$\matA$ of the network, an $n\times n$ real asymmetric matrix, where
$n$ is the number of teams (117 for Division~I-A in the 2004 season), with
element $A_{ij}$ equal to the number of times team~$j$ has beaten team~$i$
(usually 0 or 1, but occasionally~2).  The number of direct wins for a team
can be written
\begin{equation}
\mbox{direct wins for team $i$} = \sum_j A_{ji},
\end{equation}
and the number of indirect wins at distance~2 (A~beats B beats~C) as
\begin{equation}
\mbox{indirect wins at distance 2 for team $i$} = \sum_{jk} A_{kj} A_{ji},
\end{equation}
and so forth.  We discount indirect wins over direct ones by a constant
factor $\alpha$ for every level of indirection, so that an indirect win two
steps removed is discounted by~$\alpha$, an indirect win three steps
removed by~$\alpha^2$, and so forth.  The parameter~$\alpha$ will be the
single free parameter in our ranking scheme.

We now define the total \defn{win score}~$w_i$ of a team~$i$ as the sum of
direct and indirect wins at all distances thus:
\begin{eqnarray}
w_i &=& \sum_jA_{ji} + \alpha\sum_{kj}A_{kj}A_{ji} +
        \alpha^2\sum_{hkj}A_{hk}A_{kj}A_{ji}+\cdots\nonumber\\
    &=& \sum_j\bigl(1 + \alpha\sum_kA_{kj} +
        \alpha^2\sum_{hk}A_{hk}A_{kj}+\cdots\bigr)A_{ji}\nonumber\\
    &=& \sum_j(1+\alpha w_j)A_{ji}
     =  \kout_i + \alpha\sum_jA^{\mathrm{T}}_{ij}w_j,
\label{windef}
\end{eqnarray}
where $\kout_i$ is the \defn{out-degree} of vertex~$i$ in the network---the
number of edges leading away from the vertex.  When written in this
fashion, we see that the win score can also be viewed another way, as a
linear combination of the number of games a team has won (the out-degree)
and the win scores of the other teams that it beat in those games.

Similarly the \defn{loss score}~$\l_i$ of a team is
\begin{eqnarray}
l_i &=& \sum_jA_{ij} + \alpha\sum_{jk}A_{ij}A_{jk} + 
        \alpha^2\sum_{jkh}A_{ij}A_{jk}A_{kh} + \cdots\nonumber\\
    &=& \sum_jA_{ij}\bigl(1 + \alpha\sum_kA_{jk} +
        \alpha^2\sum_{kh}A_{jk}A_{kh}+\cdots\bigr)\nonumber\\
    &=& \sum_jA_{ij}(1+\alpha l_j)
     =  k^{\mathrm{in}}_i + \alpha \sum_jA_{ij}l_j.
\label{lossdef}
\end{eqnarray}

Now we define the total score for a team to be the difference $s_i=w_i-l_i$
of the win and loss scores.  Teams are then ranked on the basis of their
total score.  With this ranking scheme, a win against a strong
opponent---one with a high win score---rewards a team heavily, while a loss
against a weak opponent---one with high loss score---has the exact opposite
effect.  Thus, as discussed above, our ranking scheme automatically
incorporates the strength of schedule into the scoring.

Equations~\eref{windef} and~\eref{lossdef} can conveniently be rewritten in
vector notation, with $\textbf{w}=(w_1,w_2,\ldots)$,
$\textbf{l}=(l_1,l_2,\ldots)$,
$\mathbf{k^{\mathrm{out}}}=({\kout}_1,{\kout}_2,\ldots)$ and
$\mathbf{k^{\mathrm{in}}}=({\kin}_1,{\kin}_2,\ldots)$, giving
\begin{equation}
\vecw = \mathbf{k^{\mathrm{out}}} + \alpha \matAt\cdot\vecw,\qquad
\vecl = \mathbf{k^{\mathrm{in}}} + \alpha \matA\cdot\vecl,
\end{equation}
or, rearranging,
\begin{equation}
\vecw = \bigl(I-\alpha\matAt\bigr)^{-1}\cdot\mathbf{k}^{\mathrm{out}},\qquad
\vecl = \bigl(I-\alpha\matA\bigr)^{-1}\cdot\mathbf{k}^{\mathrm{in}}.
\label{wlsoln}
\end{equation}
These formulas are closely related to those for a well-known matrix-based
network centrality measure due to Katz and others~\cite{Katz53a,Taylor69},
and our method can be regarded as a generalization of the Katz centrality
applied to the network representation of the schedule of games.

\subsection{The parameter $\alpha$}
Before applying our method we need to choose a value for the
parameter~$\alpha$ that appears in Eqs.~\eref{windef} and~\eref{lossdef}.
A larger value of $\alpha$ places more weight on indirect wins relative to
direct ones while a smaller one places more weight on direct wins.  (For
the special case $\alpha=0$ only direct wins count at all and the total
score for a team is simply the win-loss differential.)

There are, in general, limits to the values~$\alpha$ can take.  It is
straightforward to show that the series in Eqs.~\eref{windef}
and~\eref{lossdef} converge only if $\alpha<\lmax^{-1}$, where $\lmax$ is
the largest eigenvalue of the adjacency matrix~$\matA$.  If the network is
\defn{acyclic}---has no loops of the form A beats B beats C beats~A or
longer---then the largest eigenvalue is zero (as indeed are all
eigenvalues) and hence there is no limit on the value of~$\alpha$.  This
however is an unlikely situation: there has never yet been a season for
which there were no loops in the network.  Normally therefore there is a
finite upper bound on~$\alpha$.  Historically the values of this upper
bound have been in the range 0.2 to 0.3 (Table~\ref{Lmax}), so an indirect
win cannot count for more than a fifth to a third of a direct win.
However, the number of indirect wins is in general greater the farther out
we go in the network, i.e.,~the higher the level of indirection.  This
means that the indirect wins can still make a substantial contribution to a
team's score because of their sheer number.  An $\alpha$ close to the upper
bound gives roughly equal contributions to a team's score from indirect
wins at all distances.

\begin{table}
\begin{tabular}{c|c|c}
Year & $\lmax$ & $\lmax^{-1}$\\ \hline
1998 & 3.39401 & 0.294637 \\ \hline
1999 & 4.15120 & 0.240894 \\ \hline
2000 & 3.89529 & 0.256720 \\ \hline
2001 & 3.68025 & 0.271721 \\ \hline
2002 & 4.00933 & 0.249418 \\ \hline
2003 & 3.97901 & 0.251319 \\ \hline
2004 & 3.69253 & 0.270817
\end{tabular}
\caption{Eigenvalues $\lmax$ and their inverses $\lmax^{-1}$ from regular
seasons for the years 1998--2004.}
\label{Lmax}
\end{table}

\begin{figure}
\includegraphics[width=100mm]{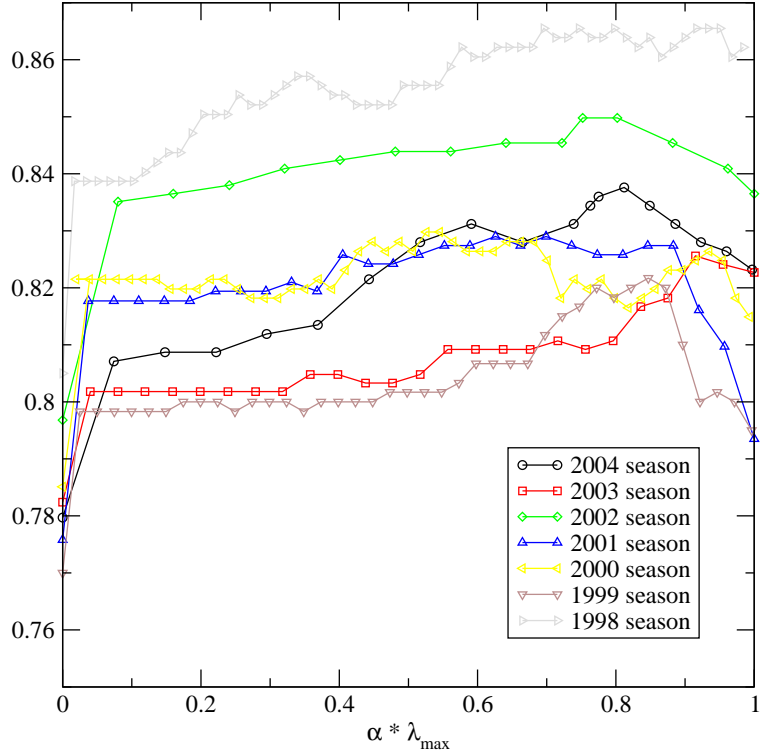}
\caption{The fraction of games won by (ultimately) higher-ranked teams in
division I-A.}
\label{WF}
\end{figure}

Aside from the limit imposed by the requirement of convergence, $\alpha$~is
essentially a free parameter, and different values will yield different
final rankings of teams.  As a simple criterion for judging which values
are best, we calculate the rankings of all teams and then examine the
\emph{retrodictive accuracy}, the fraction of all games in a season that
are won by the team with higher rank (as calculated from the final network
after the season is complete)~\cite{Martinich:2002dx}.  The results are
shown as a function of $\alpha$ for each of the years for which the BCS has
existed, 1998--2004, in Fig.~\ref{WF}.  We see that for a broad range of
values of $\alpha$ our method ranks winners above losers about 80\% of the
time---a pretty good average---and the best results appear for values of
$\alpha$ around $0.8$ of the maximum allowed value.  Thus a simple strategy
would be just to choose $\alpha=0.8\,\lmax^{-1}$.

While this strategy appears to give good results in practice, it has one
problem, namely that the calculation of~$\lmax$ (and hence of~$\alpha$)
requires a knowledge of the entire directed network, which means that we
can only perform the calculation after the end of a season once the outcome
of every game has been decided.  In practice, however, one often wants to
rank teams before the end of the season, for instance to calculate weekly
standings from partial results as the season progresses.  Thus we would
like to be able to decide the value of $\alpha$ before the start of the
season.  In the next section we provide a method for doing this that
appears to work well.

\subsection{An algorithm for choosing $\alpha$}
\label{alphaalg}
As discussed in the preceding section, the limit $\lmax^{-1}$ on $\alpha$
would be infinite were there no loops in the network.  Only if there are
loops (which there usually are) does the limit become finite.  And in
general the more loops there are the lower the limit.  (The combinatorial
explosion in the combinations of loops that paths can circulate around
makes the number of paths increase faster with path length when there are
more loops, and this then requires a lower value of~$\alpha$ to ensure
convergence.)  Real schedule networks have fewer loops than one would
expect on the basis of chance, precisely because teams do vary in strength
which gives rise to networks that are close to being acyclic.  Thus we
would expect the value of $\lmax$ to be lower and the limit $\lmax^{-1}$ to
be higher in a real network than in a network with randomly assigned edges.
(As a check, we have performed Monte Carlo randomizations of the edge
directions in the real networks and find that indeed $\lmax$ consistently
increases when we do this.)  This provides us with way to calculate a safe
and practical upper bound on the value of $\alpha$ without knowledge of
game outcomes: we simply calculate the limit for a network with randomly
assigned outcomes.

It is straightforward to calculate the largest eigenvalue for a random
directed network in which the distribution of in- and out-degrees is known.
Let $P(\kin=i,\kout=j)$ be the joint probability distribution of in- and
out-degrees.  The largest eigenvalue is equal to the factor by which the
number of paths of a given length starting at a vertex increases when we
increase that length by one, in the limit of long path length.  But this is
simply equal to the mean out-degree of the vertex reached by following a
randomly chosen edge in the graph, which is given by~\cite{NSW01}
\begin{equation}
{\sum_{ij} ij P(\kin=i,\kout=j)\over\sum_{ij} iP(\kin=i,\kout=j)}
  = {\av{\kin\kout}\over\av{\kin}}.
\end{equation}
For our random network, the joint degree distribution is derived by
randomly assigning directions to edges on an initially undirected network
whose degree distribution is given by the distribution of the number of
games played by the teams in the regular season.  Let us denote this
distribution~$p_k$.  Then
\begin{equation}
P(\kin=i,\kout=j) =  2^{-(i+j)} {i+j\choose i} p_{i+j},
\end{equation}
and so our expression for the largest eigenvalue is
\begin{eqnarray}
\lmax &=& {\sum_{i,j=0}^\infty ij  2^{-(i+j)} {i+j\choose i} p_{i+j}\over
       \sum_{i,j=0}^\infty i 2^{-(i+j)} {i+j\choose i} p_{i+j}}
   =  {\sum_{i=0}^\infty \sum_{k=i}^\infty i(k-i) 2^{-k} {k\choose i}
       p_k\over
       \sum_{i=0}^\infty \sum_{k=i}^\infty i 2^{-k} {k\choose i} p_k}
      \nonumber\\
  &=& {\sum_{k=0}^\infty 2^{-k} p_k \sum_{i=0}^k i(k-i) {k\choose i}\over
       \sum_{k=0}^\infty 2^{-k} p_k \sum_{i=0}^k i {k\choose i}}
   =  {\sum_{k=0}^\infty 2^{-k} p_k k(k-1) 2^{k-2}\over
       \sum_{k=0}^\infty 2^{-k} p_k k 2^{k-1}}\nonumber\\
  &=& {\av{k^2} - \av{k}\over2\av{k}},
\label{rglmax}
\end{eqnarray}
where $\av{k}$ and $\av{k^2}$ are the mean and mean-square number of games
played by a team in the season of interest.

As a test of this calculation we have calculated numerically the actual
values of $\lmax$ for simulated seasons with randomly assigned wins and
losses.  The results are shown in Table~\ref{randomlmax}.  As the table
shows, agreement between the analytic calculation and the simulations is
excellent.

\begin{table}
\begin{tabular}{c|c|c}
Year & $\lmax$ from MC  simulation & $(\av{k^2}-\av{k})/2\av{k}$ \\ \hline
1998 & $4.957\pm0.068$ & $4.935$ \\
1999 & $4.901\pm0.066$ & $4.882$ \\
2000 & $4.927\pm0.066$ & $4.896$ \\
2001 & $4.896\pm0.065$ & $4.875$ \\
2002 & $5.350\pm0.069$ & $5.334$ \\
2003 & $5.277\pm0.064$ & $5.260$ \\
2004 & $4.859\pm0.065$ & $4.838$
\end{tabular}
\caption{Comparison of the values of $\lmax$ calculated from Monte Carlo
simulations and using Eq.~\eref{rglmax} for the years 1998--2004.}
\label{randomlmax}
\end{table}

All the values of $\lmax$ in Table~\ref{randomlmax} are larger by about
20\% than the actual $\lmax$ for the corresponding season
(Table~\ref{Lmax}), precisely because actual wins and losses are not
random, but reflect the real strengths and weaknesses of the teams.  But
this means that the random-graph value of $\lmax$ imposes a limit on
$\alpha$ that will in general be about $0.8$ of the limit derived from the
true final schedule network incorporating the real wins and losses.  And
this value is right in the middle of the region found in the preceding section to give the best rankings of the teams.  Thus an elegant solution
to the problem of choosing $\alpha$ emerges.  We simply choose a value
equal to the limiting value set by Eq.~\eref{rglmax}:
\begin{equation}
\alpha = {2\av{k}\over\av{k^2}-\av{k}}.
\label{alphadef}
\end{equation}
This guarantees convergence, requires no knowledge of the eventual outcome
of games, and appears to give optimal or near-optimal rankings under these
constraints.  This then is the value that we will use in the calculations
in this paper.

\subsection{Comparison with the BCS rankings}
\label{comparison}
We now compare the results of our method with the official BCS ranking
results.  It is worth pointing out that agreement with the official
rankings is not necessarily a sign of success for our method.  If our
method were better in some sense than the official method (for example in
comparison with the opinions of human expert judges), then necessarily the
two methods would have to disagree on some results.  Nonetheless, since the
BCS ranking is, by common consent, officially the collective wisdom, it is
clearly of interest to see how our method compares with it.

First, in Table~\ref{RS2004} we show the rankings calculated from our
method and from the BCS computer algorithms for the top 25 BCS teams of
2004.  The value of $\alpha$ for 2004 from Eq.~\eref{alphadef} is $0.207$,
or about $0.763$ of the maximum.

\begin{table}
\begin{tabular}{c|c|c|c}
BCS & School & Our method & BCS Computers\\
\hline
1 & Southern California &  2  &  2\\
2 & Oklahoma            &  1  &  1\\
3 & Auburn              &  3  &  3\\
4 & Texas               &  4  &  4\\
5 & California          &  8  &  6\\
6 & Utah                &  5  &  5\\
7 & Georgia             & 16  &  8\\
8 & Virginia Tech       &  6  &  T-9\\
9 & Boise State         &  7  &  7\\
10 & Louisville          & 11  & 13\\
11 & Louisiana State     & 15  & T-9\\
12 & Iowa                & 10  & 12\\
13 & Michigan            & 14  & 17\\
14 & Miami (FL)          &  9  & T-14\\
15 & Tennessee           & 17  & T-14\\
16 & Florida State       & 12  & 21\\
17 & Wisconsin           & 20  & 20\\
18 & Virginia            & 18  & 18\\
19 & Arizona St.         & 13  & 11\\
20 & Texas A\&M          & 19  & 16\\
21 & Pittsburgh          & 27  & --\\
22 & Texas Tech          & 23  & 22\\
23 & Florida             & 26  & --\\
24 & Oklahoma State      & 21  & 19\\
25 & Ohio State          & 22  & --\\
\end{tabular}
\caption{Comparison of standings for the final top 25 BCS teams in 2004, calculated using the method described in this paper and the standard BCS composite computer ranking. ``--''denotes a team that was not ranked among the top 25 in the BCS composite computer ranking and ``T'' denotes a tied rank.}
\label{RS2004}
\end{table}

Even from a casual inspection it is clear that there is a reasonable match
between our rankings and the official ones.  For instance, the correlation
coefficient between the two sets of rankings is $0.90$.  Given the
simplicity of our method, it is pleasantly surprising that the rankings are
in such good agreement with other far more complicated algorithms.

Among these 25 teams, our method classified two---Pittsburgh and
Florida---to be outside the top 25.  Interestingly, these same teams were
also ranked outside the top 25 by all the BCS computer algorithms except
the Billingsley algorithm.  Furthermore, none of the other computer polls
does any better at predicting the final top 25---each gets at least two
wrong.  Other points of interest are the ranks of the Universities of
Auburn, Texas, and California.  Auburn, although undefeated in the regular
season, did not participate in a championship game because it was
consistently ranked third in all polls, and our method concurs.  Texas and
California played very similar seasons but both human polls ranked
California to be higher, while all the computer polls said the reverse.
Our method lines up with the computer polls in this respect.

In Table~\ref{RS2004onefive}, we compare the top five BCS teams for each
year (with $\alpha$ again selected as described in Section~\ref{alphaalg}
and taking values typically between $0.7$ and $0.85$ of the maximum
allowable value).  The rankings consistently agree on at least three of the
top five teams in each year.

\begin{table}
\begin{tabular}{|c|c|c|c|c|c|}\hline
\multicolumn{2}{|c|}{2003}     & \multicolumn{2}{|c|}{2002}  &  \multicolumn{2}{|c|}{2001} 
\\ \hline
Our method     & BCS           & Our method   & BCS          &    Our method     & BCS \\ \hline
Oklahoma       & Oklahoma      & Ohio St.     & Miami (FL)   &    Tennessee (6)  & Miami (FL) \\
Southern Cal   & Louisiana St. & Southern Cal & Ohio St.     &    Miami (FL)     & Nebraska \\ 
Florid St. (7) & Southern Cal  & Miami (FL)   & Georgia      &    Illinois (8)   & Colorado \\
Louisiana St   & Michigan (10) & Georgia      & Southern Cal &  Colorado         & Oregon (6)\\
Miami (FL) (9) & Ohio St. (6)  & Oklahoma (7) & Iowa (8)     &  Nebraska         & Florida (7) \\ \hline\hline
\multicolumn{2}{|c|}{2000} & \multicolumn{2}{|c|}{1999}      & \multicolumn{2}{|c|}{1998} 
\\ \hline
Our method     & BCS           & Our method   & BCS          & Our method     & BCS \\ \hline
Washington     & Oklahoma      & Florida St.  & Florida St.  & UCLA           & Tennessee  \\
Oklahoma       & Florida St.   & Mich. St. (9)& VA Tech (6)  & Florida St.    & Florida St \\
Oregon St. (6) & Miami (FL) (8)& Nebraska     & Nebraska     & Texas A\&M (6) & Kansas St. \\
Florida St.    & Washington    & Michigan  (8)& Alabama      & Tennessee      & Ohio St. (7)\\
Oregon (10)    & VA Tech (15)  & Alabama      & Tennessee (8)& Kansas St.     & UCLA     \\ \hline
\end{tabular}
\caption{Comparison of the top five teams calculated using the method
presented in this paper and using the complete BCS composite ranking
(including human polls) for the years 1998--2003.  Numbers in parentheses
for our method denote teams' ranks under BCS, and \emph{vice versa}.}
\label{RS2004onefive}
\end{table}

For a more quantitative comparison of our method with the official BCS
rankings we have also examined the retrodictive accuracy, as defined
earlier.  Since the BCS announces only the final top $15$ or $25$ teams in
its rankings each year, we have constrained our calculation of the
retrodictive accuracy to the games played among those teams.  The results
are given in Table~\ref{Retrodictive}.

\begin{table}
\begin{tabular}{c|c|c|c|c|c|c|c}
           & 2004  & 2003  & 2002  & 2001 & 2000  & 1999  & 1998 \\ \hline
Our method & 0.81  & 0.62  & 0.89  & 0.47 & 0.65  & 0.67  & 0.83 \\ 
           & 35/43 & 23/37 & 16/18 & 7/15 & 15/23 & 8/12  & 15/18 \\ \hline
BCS        & 0.84  & 0.69  & 0.79  & 0.47 & 0.75  & 0.69  & 0.83 \\
           & 29/37 & 25/36 & 15/19 & 8/17 & 12/16 & 11/16 & 10/12 \\ \hline
\end{tabular}
\caption{Retrodictive accuracies and the numbers of games played among the
top $25$ (2004-2003) or $15$ (2002-1998) teams for our method and for the
official BCS rankings.}
\label{Retrodictive}
\end{table}

\section{A non-football example application}
The discussions so far in this paper have focused on the problem of
ranking teams in American college football, but our method could in
principle be used for a wide variety of other ranking problems in which
individuals, groups, or objects are compared in a pairwise fashion.  Such
problems come up in very many fields of scientific interest.  Here we give
one example from biology, of dominance hierarchies among animals.

In a study published in 1979, Lott~\cite{Lott:1979hg} observed the pattern
of dominant and submissive behaviors between 26 male American bison in
Montana over a period of about a month.  Pairs of bison engaged in
aggressive interactions to establish dominance within the herd and the
outcome of each observed interaction was noted, creating a directed network
of ``wins'' and ``losses'' between pairs of animals, just as in the college
football case.  Lott also observed the breeding success of the bison over
the same period, as measured by the number of mating events between the
male bison involved in the dominance hierarchy and the cows (who do not
engage in the aggressive interactions).

We used our method to calculate a ranking for the bison, using a value
of $\alpha=0.124$ from Eq.~\eref{alphadef}.  We find that the
resulting rankings are highly correlated with breeding success of the
bison, with a correlation coefficient of $0.611$.  Correlation between
status and breeding success is a widely understood feature of breeding
populations~\cite{Cole:1981hy,Lott:1979hg}; our method provides a
quantitative confirmation in this particular case, and could in principle
be applied to other examples for which similar data are available.

\begin{figure}
\includegraphics[width=100mm]{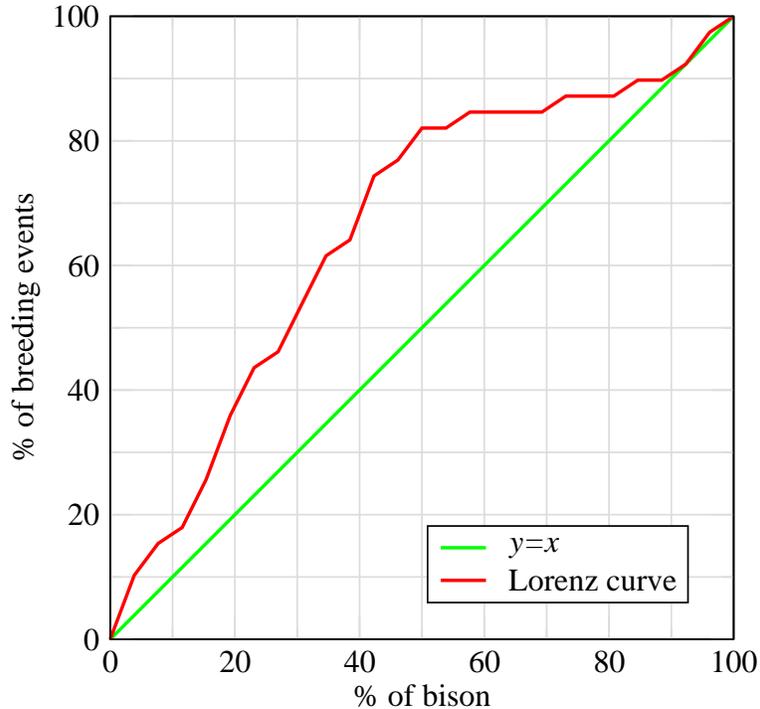}
\caption{A Lorenz curve (red) of breeding events versus the rankings of the
bison.  It is placed consistently above the equality line (green) which
represents a perfectly even distribution, indicating a positive correlation
between breeding power and dominance hierarchy.}
\label{bisonlorenz}
\end{figure}

One can also represent the correlation between rank and breeding success
using a so-called Lorenz curve---a device often used to represent inequality in the distribution of wealth between the richest and poorest
individuals.  We show a Lorenz curve for our bison example in
Fig.~\ref{bisonlorenz}.  The curve is a plot of the fraction of mating
events that involve bison of a given rank or higher, as a function of
rank.  If mating success were independent of rank, the curve would follow
the $45^\circ$ line, but instead it deviates markedly from the line and the
size of this deviation is a measure of the extent to which higher ranked
bison are more successful in mating.  (The area between the $45^\circ$ line
and the curve is called the Gini coefficient, and is sometimes quoted as a
measure of inequality, although in this case we feel the correlation
coefficient of rank and mating success is a more easily understood measure
of the extent to which our rankings predict success.)

\section{Conclusions}
In this paper we have introduced a ranking system for division I-A American
college football based on a common-sense argument frequently made by sports
fans (and not limited to American football).  The method has an elegant
mathematical formulation in terms of networks and linear algebra that is
related to a well-known centrality measure for networks.  The method has
one free parameter and we have given empirical evidence indicating the
typical range of best values for this parameter and a method for choosing
a value in any particular case.

Applying our method to the seven years during which the BCS ranking scheme
has existed, we find excellent agreement between the method and the
official rankings but with some deviations, particularly in well-known
controversial cases.  We believe that the combination of sound and
believable results with a strong common-sense foundation makes our method
an attractive ranking scheme for college football.  Our method can be
applied to other ranking problems outside of college football as well,
including other games or sports, or other problems entirely.  We have given
one example application to a dominance hierarchy in a herd of American bison.

Finally, we would like to comment on the mathematical generalizability of
our method.  The method lends itself readily to the addition of other
elements, such as margin of victory, home-field advantage, progress of the
season, and so forth: these could be introduced as modifiers on the weights
of the edges in the network, and it would be interesting to see how these
affect the outcome of the method.  However, we believe that the very
simplicity of the current method, with its single parameter, is a
substantial advantage, and that simplicity should be encouraged in these
types of mathematical methods.  A method such as ours reduces the extent to
which the calculations must be tuned to give good results while at the same
time retaining an intuitive foundation and mathematical clarity that makes
the results persuasive.

\begin{acknowledgments}
The authors would like to thank Elizabeth Leicht for useful comments.  This
work was funded in part by the National Science Foundation under grant
number DMS--0405348 and by the James S. McDonnell Foundation.
\end{acknowledgments}

\bibliographystyle{siam}
\bibliography{references}

\end{document}